\documentclass[3p,times,twocolumn]{elsarticle}

\usepackage{epsfig}
\usepackage{graphicx}
\usepackage{epstopdf}

\usepackage{amssymb}

\usepackage{lineno}

\usepackage[figuresright]{rotating}

\usepackage{amssymb}
\usepackage{subfigure}

\begin{document}
%
\begin{frontmatter}

\title{
Measurement of basic characteristics and gain uniformity of a triple GEM detector}

\author[a]{Rajendra~Nath~Patra}
\author[a]{Rama~N.~Singaraju}
\author[b]{Saikat~Biswas}
\author[a]{Zubayer~Ahammed}
\author[a,c]{Tapan~K.~Nayak}
\author[a]{Yogendra~P.~Viyogi}

\address[a]{Variable Energy Cyclotron Centre, HBNI, Kolkata-700064, India}
\address[b]{Bose Institute, Department of Physics and CAPSS, Kolkata-700091,	India}
\address[c]{CERN, Geneva~23, Switzerland}

\begin{abstract}

Large area Gas Electron Multiplier (GEM) detectors have been the preferred choice for tracking
devices in major nuclear and particle physics experiments. Uniformity 
over surface of the detector in terms of gain, energy resolution and efficiency is
crucial for the optimum performance of these detectors. 
In the present work, detailed performance study of a 10$\times$10 cm$^{2}$
triple GEM detector operated using Ar and CO$_{2}$ gas mixtures
in proportions of 70:30 and 90:10,
has been made by making a voltage scan of the
efficiency with $^{106}$Ru-Rh $\beta$-source and cosmic rays. The gain
and energy resolution of the detector were studied using the X-ray
spectrum of $^{55}$Fe source. The uniformity of the detector 
has been investigated by dividing the detector in 7$\times$7 zones and measuring the gain and energy resolution at the center of each zone. The variations of the gain and energy resolution have been found to be 8.8\% and 6.7\%, respectively. These studies are essential to characterise
GEM detectors before their final use in the experiments.

\end{abstract}

\end{frontmatter}

\section{Introduction }\label{intro}

New generation nuclear and particle physics experiments require
charged particle tracking devices with low material budget and
excellent position resolutions. For the last several decades, various
types of micro-pattern gas detectors (MPGD) have been developed for their use
in experiments at major accelerator facilities as well as for
applications in imaging technologies. Gas Electron Multiplier (GEM) 
detectors, developed at CERN in 1997~\cite{sauli,sauli2016}, is one of the 
new generation MPGD, which fulfils the stringent conditions of
existing and proposed large scale experiments. With increasing
energy and beam luminosity in accelerator facilities, the requirements
for detector technologies have been continuously changing. Because of their 
high rate capability, fast timing, good position 
resolution and ion suppression 
 features, GEM detectors have widely been chosen as preferred tracking devices 
 in particle and heavy-ion physics experiments~\cite{benci,alfonsi,ball}. 
Charged particle tracking devices in experiments at
Brookhaven National Laboratory~\cite{phenix1,huang}, 
CERN Large Hadron Collider~\cite{alicetpc, alicetpc1, cmsgem, cmsgem1},
the Facility for Antiproton and Ion Research (FAIR) at
GSI~\cite{cbmgem, saikat, cbmgem1} and future experiments in International Linear Collider (ILC)~\cite{ilc} have chosen GEM detectors.

A GEM foil consists of an insulator made of a 50~$\mu$m thick Kapton foil with 5~$\mu$m thick copper cladding on both sides and pierced by a regular array of holes. These perforated holes, having typically 70~$\mu$m diameter and separated by 140~$\mu$m pitch, are arranged in a hexagonal pattern. Depending on the etching method, the holes are single conical or bi-conical. 
A bias voltage of 350-400~V applied across the GEM foils, provides a very high electric field ($\sim$70~kV/cm) in the
holes because the field lines are focused in the holes. This creates
large gas amplification, up to several thousands. 
The electrons are collected on an anode, which is the readout
plane. Due to the micro-hole structure, excellent spatial resolution
(of the order of 100~$\mu$m)
along with a reasonable time resolution (about 10~ns) can be achieved.
Another major advantage of the GEM detectors over other
MPGD designs is its ion back flow (IBF) suppression~\cite{ball} and low discharge
probability~\cite{croci}. In a multistack GEM detector setup, the hole diameter,
pitch, electric field across different gas gap can be optimised to minimise the IBF. Discharge probability is reduced as the total electron multiplication is divided into many steps
across the holes of the different GEM foils. These features make the
GEM detectors suitable candidates for high energy physics experiments.

The GEM detectors, like any other gas detectors, can be characterised
in terms of gain, efficiency of detection of charged particles, energy
resolution, time resolution, position resolution and rate
capability. For a given detector configuration, these parameters vary
with applied electric field, gas pressure and temperature. The optimum
running conditions are determined depending on the use of the
detector. For effective tracking of charged particles, it is essential
to have uniform gain over the entire area of the GEM detector. The
gain spatial uniformity depends on the quality of GEM foils,
uniformity of the holes, fabrication steps and other such quality
control parameters. Thus the measurement of gain uniformity forms a
basic quality assurance, especially in the case of large area detectors.

In the present study, we have first measured the 
basic characteristics of a standard triple GEM detector
10$\times$10 cm$^{2}$ in size. The design of the GEM detector and test setup are presented in
the next section. The detector responses to $\beta$-source and
cosmic rays are discussed in section~3. Detector characteristics in
terms of efficiency, gain, energy and time resolutions are presented
in section~4. 
We have adopted a new method to measure spatial variations of gain and
energy resolution. The respective
results are presented and discussed in section~5. A summary and outlook are given in section~6.
\section{Detector design and experimental setup}
\begin{figure}[!ht]
\begin{center}
\includegraphics[scale = 0.4]{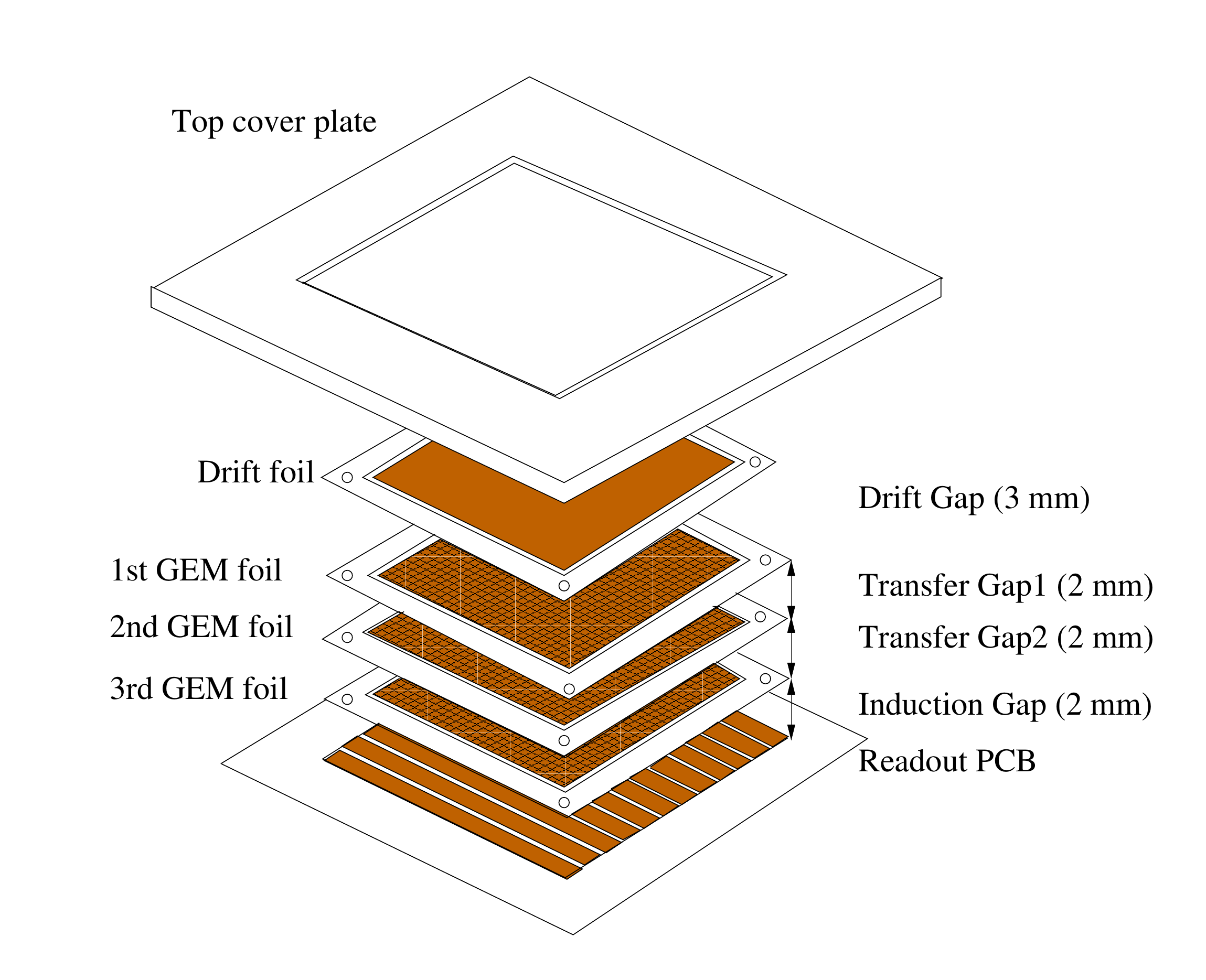}
\caption[short]{Geometrical design of the triple GEM detector.}
\label{GEM_layout}
\end{center}
\end{figure}

The layout of a triple GEM detector is shown in Fig.~\ref{GEM_layout}.
The detector is constructed by stacking three standard single mask 
stretched GEM foils manufactured at CERN. Each foil has 10$\times$10
cm$^{2}$ area, having 70~$\mu$m diameter holes arranged in a 140~$\mu$m pitch network.
In a multi stack GEM detector, the drift,
multiplication and induction regions are kept physically separated as
shown in the figure. Thus, there is a freedom to design the readout
according to the requirement of the experiment. 
In the present setup, the drift gap, transfer gaps and induction gap are kept as 3-2-2-2~mm, 
respectively. The drift plane is made of 
a Kapton foil cladded on one side with a thin (5~$\mu$m) layer of copper
and the entire detector is
kept in a gas tight box. In our setup, the detector has been operated using Argon and CO$_2$
gas mixtures in proportions of 70:30 and 90:10 at atmospheric
pressure.
\begin{figure}[!ht]
\begin{center}
\includegraphics[scale=0.14]{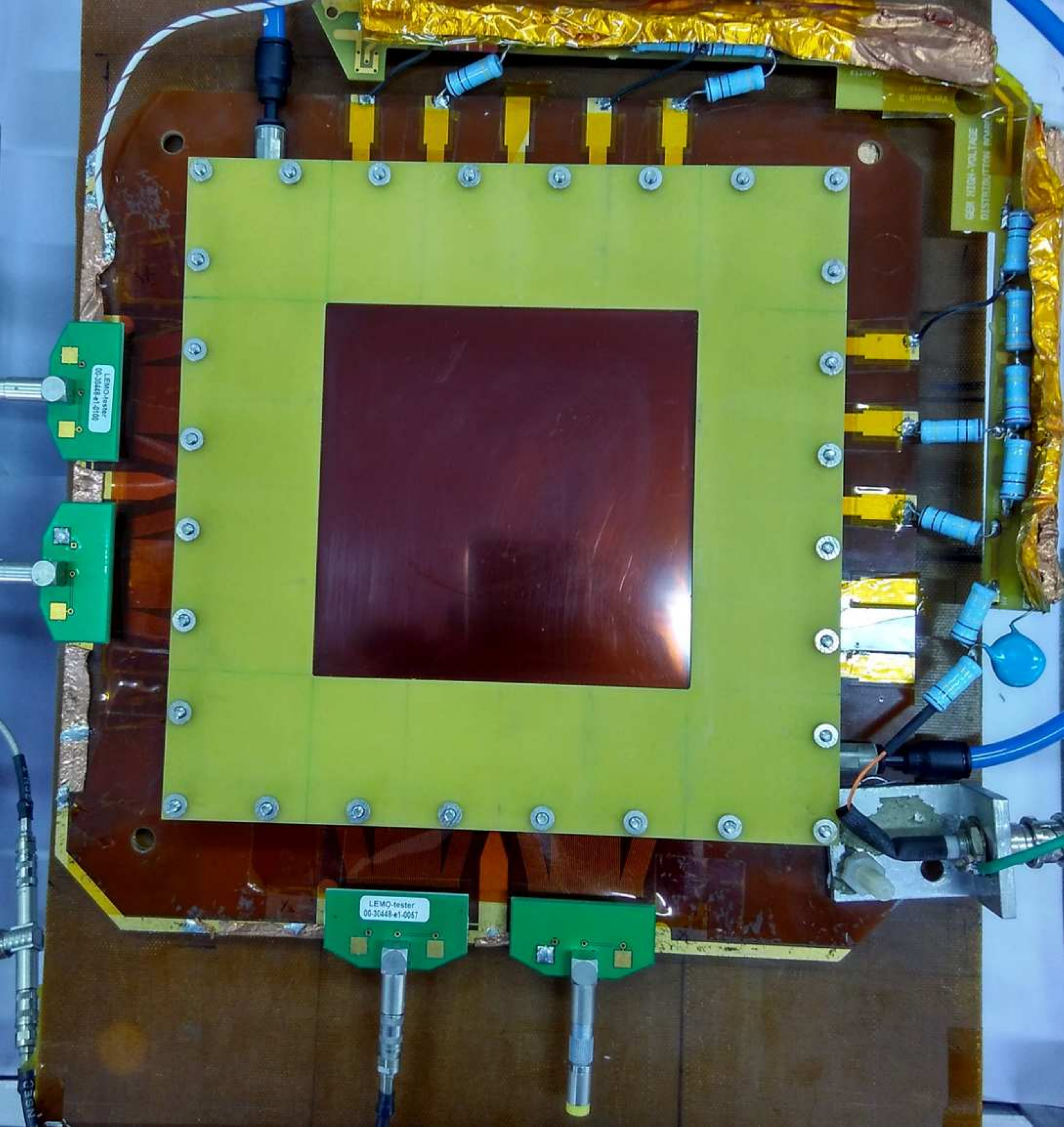}
\caption[short]{Photograph of the laboratory 
  setup of the triple GEM detector.}
\label{GEM_photo}
\end{center}
\end{figure}
A voltage difference is applied through a
voltage divider resistor chain to produce the required electric field
across the GEM foils and in the gas gaps. The resistor chain contains 
a filter circuit for noise reduction. The design of the resistor chain
is made in such a way that a negative high voltage (HV) within the
range of 3900-4500~V can be applied. A photograph of the complete
triple GEM detector setup along with the voltage divider is shown in
Fig.~\ref{GEM_photo}.

Detector tests are performed by varying the HV. With application of
this voltage difference, primary electrons are produced inside the
drift gap in presence of incident radiation
and those electrons drift towards the first GEM foil.
After three stages of charge multiplications within the holes of the
GEM foils, the 
gain of the detector reaches values as high as 10$^{3}$-10$^{5}$.
All electrons drifting to the induction gap are collected in the
readout plane. The readout of the detector consists of a base plate with 256 X- and
256 Y- metallic strips. Each of the 256 strips is connected to two
128 pin connectors. A sum-up board adds up signals from 128 readout
strips. In total, 4 sum-up boards provide the measured signals.

\section{Detector response}
\begin{figure}[!ht]
\begin{center}
\includegraphics[scale = 0.4]{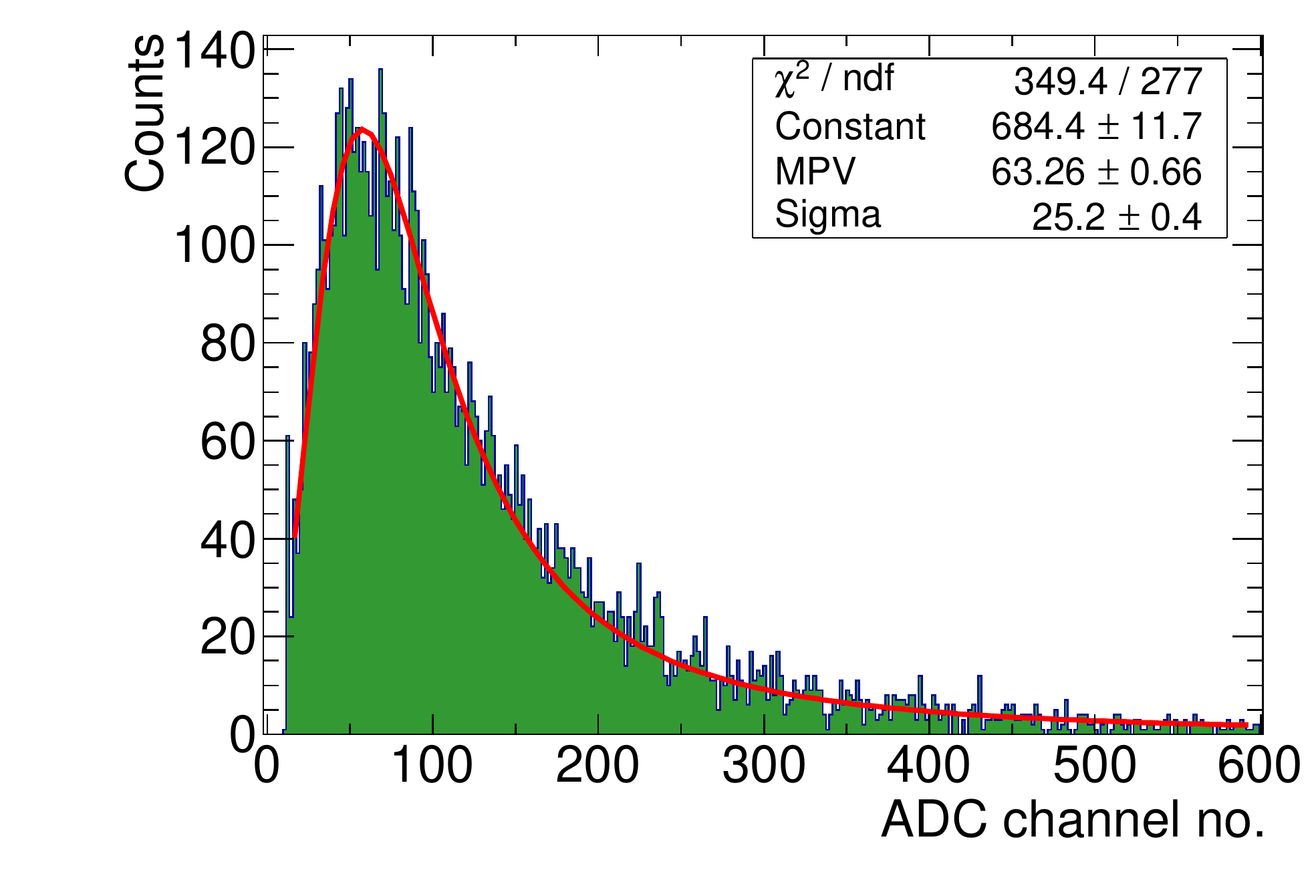}
\includegraphics[scale = 0.4]{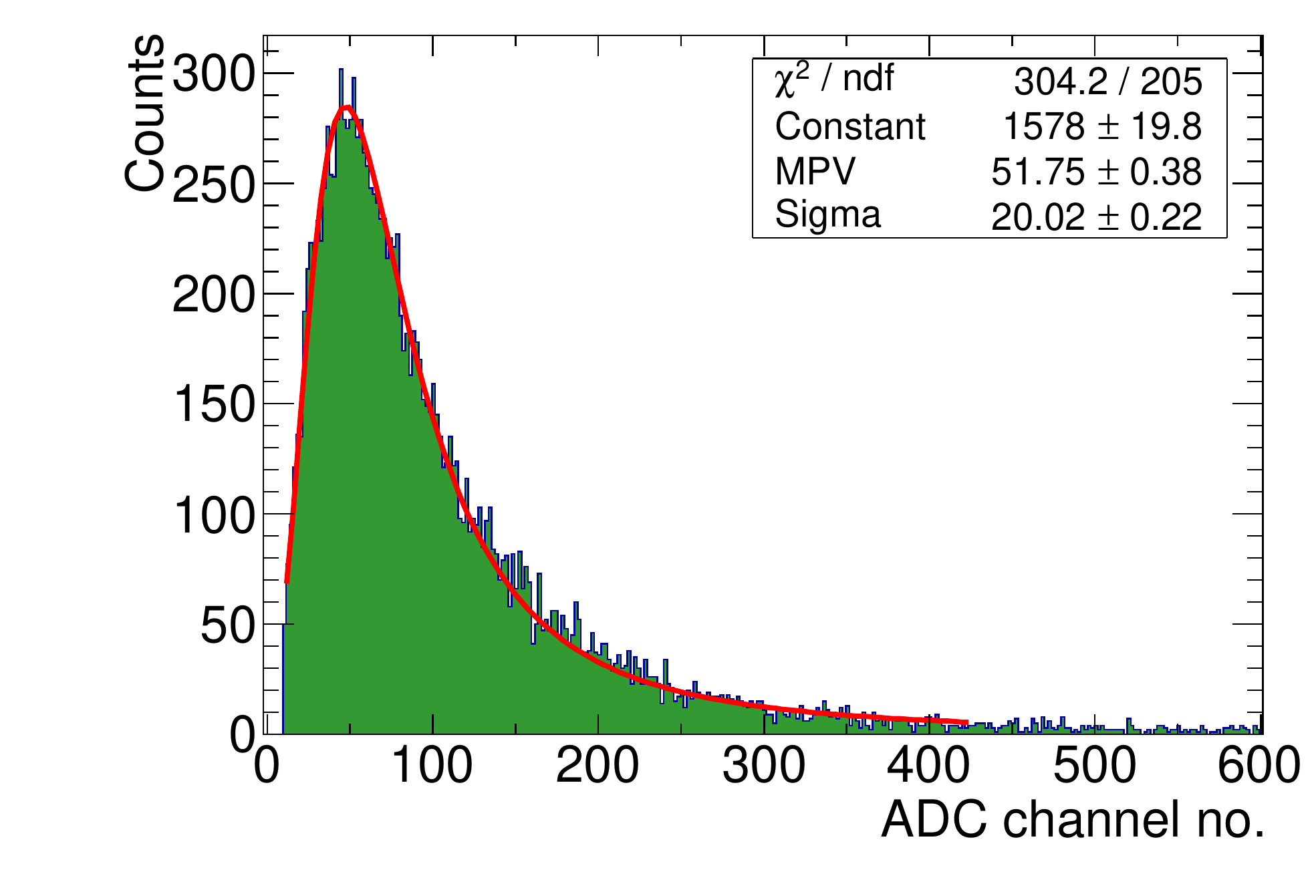}
\caption[short]{(top) cosmic muon and (bottom)
$^{106}$Ru-Rh $\beta$-spectrum at 4400 V for the GEM detector.}
\label{cosmic-ru}
\end{center}
\end{figure}

The response of the GEM detector has been studied after
stabilisation of the gas flow and after minimising the electronics
noise~\cite{saikat, rajendra}. The first test of the detector has been performed with cosmic muon. The
trigger setup for this study consisted of two cross scintillators
placed above the detector and a third scintillator below.
A valid trigger consists of coincidence of the signals from all three
scintillators. Cosmic muon spectra with gated
trigger have been obtained with different HV settings. 
The top panel of Fig.~\ref{cosmic-ru}
shows the cosmic muon ADC spectrum taken at 4400 V. 
The spectrum is fitted with a Landau distribution to have the Most
Probable Value (MPV). Studies with cosmic muons take a long time and
a faster way to obtain the efficiency is to use a $^{106}$Ru-Rh
$\beta$-source. The trigger setup of three scintillators remains the
same as used for cosmic muons. The ADC spectrum corresponding to
$^{106}$Ru-Rh $\beta$-source at 4400 V, is shown in the
bottom panel of Fig.~\ref{cosmic-ru}. In the $\beta$-spectrum the MPV
value is a bit lower compared to that of cosmic rays. 

\begin{figure}[!ht]
\begin{center}
\includegraphics[scale = 0.4]{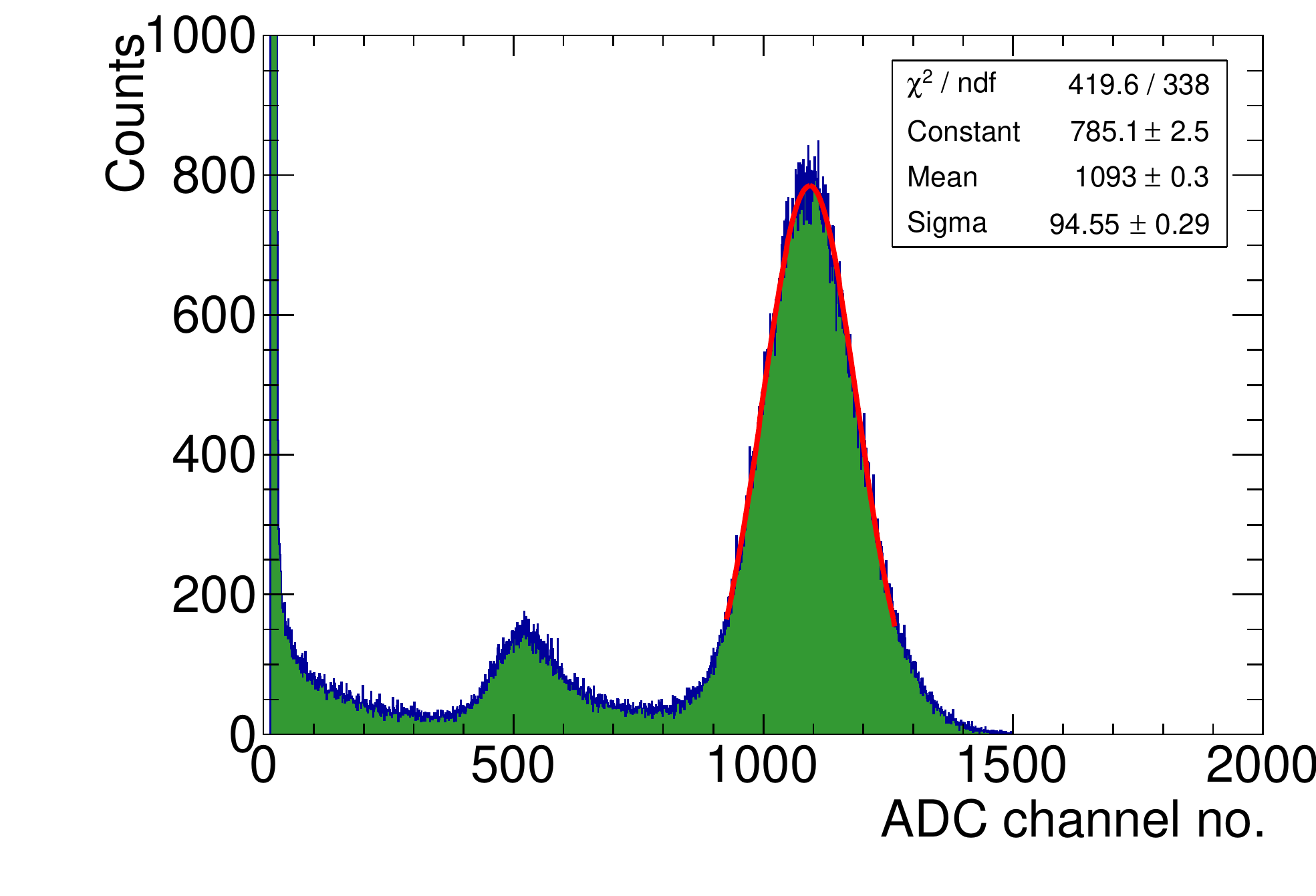}
\caption[short]{$^{55}$Fe spectrum at 4400V for the GEM detector.}
\label{fe}
\end{center}
\end{figure}

In order to study the detector gain and energy resolution, the
detector has been tested with a $^{55}$Fe X-ray source which provides
5.9~keV X-rays. The pulse height spectrum at 4400~V for this source is
shown in Fig.~\ref{fe}. The peak at the higher ADC channel
corresponds to the 5.9~keV main photo peak and the one to the left 
is identified as the Argon escape peak. 

\section{Detector characteristics}
Detailed characteristic studies, such as, detector efficiency, gain, energy resolution and
time resolution have been performed 
for the GEM detector using cosmic rays and different radioactive sources. Results of these studies are presented and discussed below.
\subsection{Detector efficiency}
Efficiency of the GEM detector has been studied with cosmic rays and
$^{106}$Ru-Rh $\beta$-source. For this measurement, trigger was provided by the coincidence signal of a set of three scintillators, as described in the previous section. Number of triggered particles
giving signal on the GEM detector yields the efficiency. 

\begin{figure}[!ht]
\raggedleft
\includegraphics[scale = 0.4]{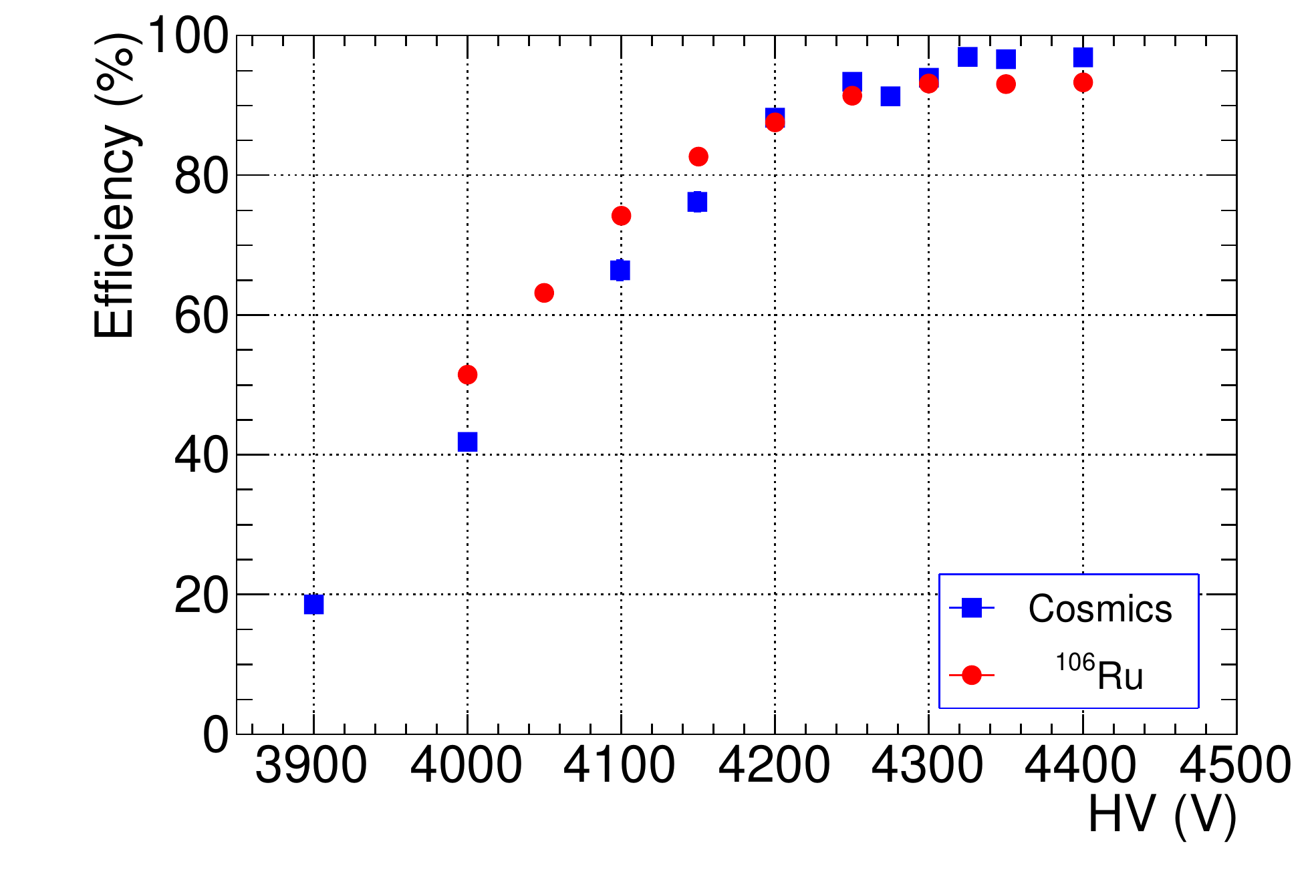}
\includegraphics[scale = 0.4]{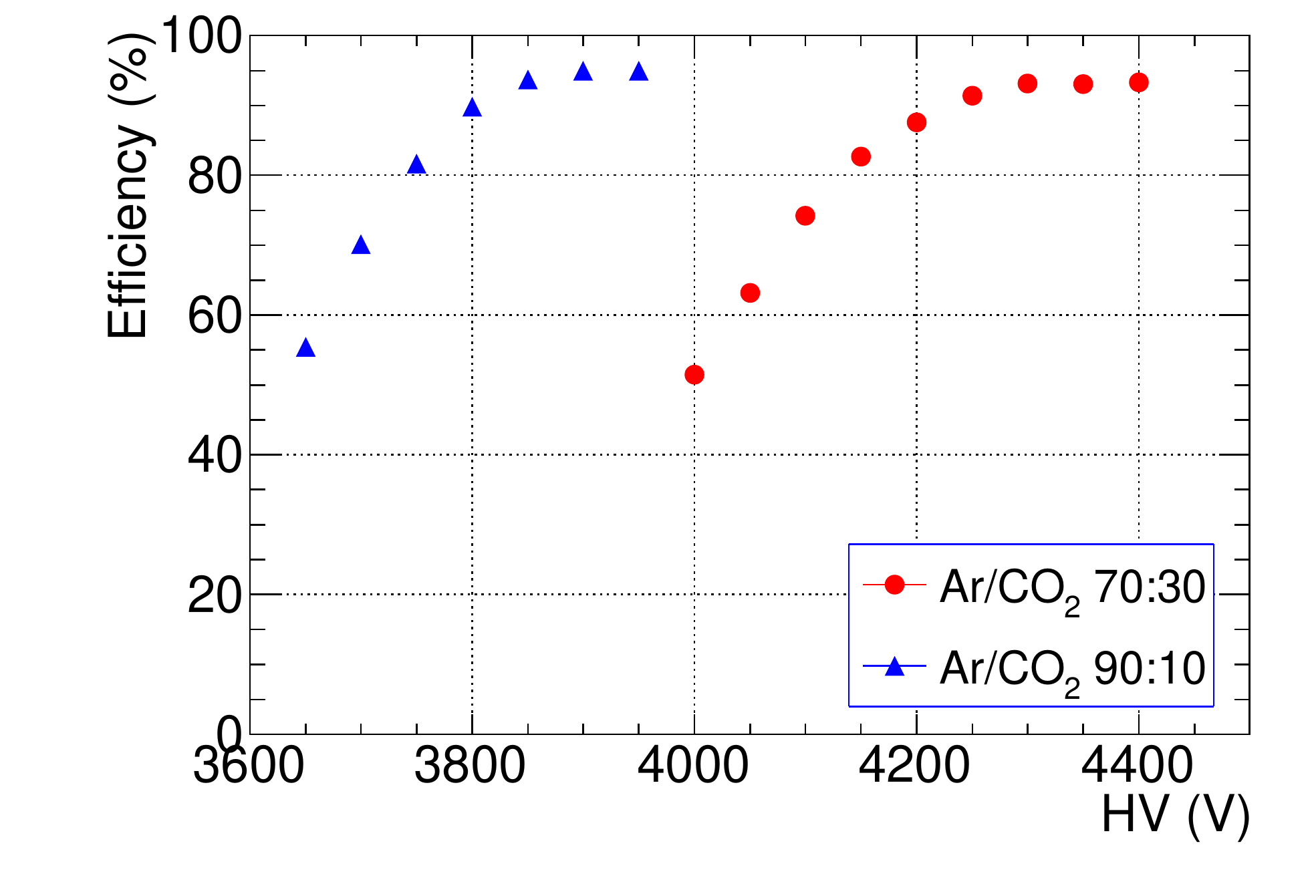}
\caption[short]{(top) Efficiency plot for the GEM detector, as a function of HV obtained by using cosmic muon and 
 $^{106}$Ru-Rh source with Ar-CO$_2$ 70:30 gas and (bottom) efficiency plot with different Ar-CO$_2$ gas mixtures, using the  $^{106}$Ru-Rh $\beta$-source.}
\label{eff}
\end{figure}

Efficiency obtained as a function of applied HV is shown in Fig.~\ref{eff}.
The top panel gives the efficiency for Ar-CO$_2$ gas mixture in
70:30 proportions for cosmic rays and $^{106}$Ru-Rh source. 
The results are very similar. Increasing the
HV from 3900V, the efficiency increases from 20\% and beyond 4300V the
efficiency comes to a plateau region  at $\sim$95\% level. 
The bottom panel of Fig.~\ref{eff} shows the efficiency as a function
of HV for Ar-CO$_2$ gas mixture in two proportions, 70:30 and
90:10. These measurements were performed by using the $^{106}$Ru-Rh $\beta$-source. Similar efficiency values can be achieved for 90:10 gas mixture at a much lower voltage compared to that of 70:30. The optimum detector operation for the 70:30 gas mixture is around 4300V, whereas for the 90:10 mixture, this voltage is about 3850V.
\subsection{Detector gain}
\begin{figure}[!ht]
\begin{center}
\includegraphics[scale = 0.4]{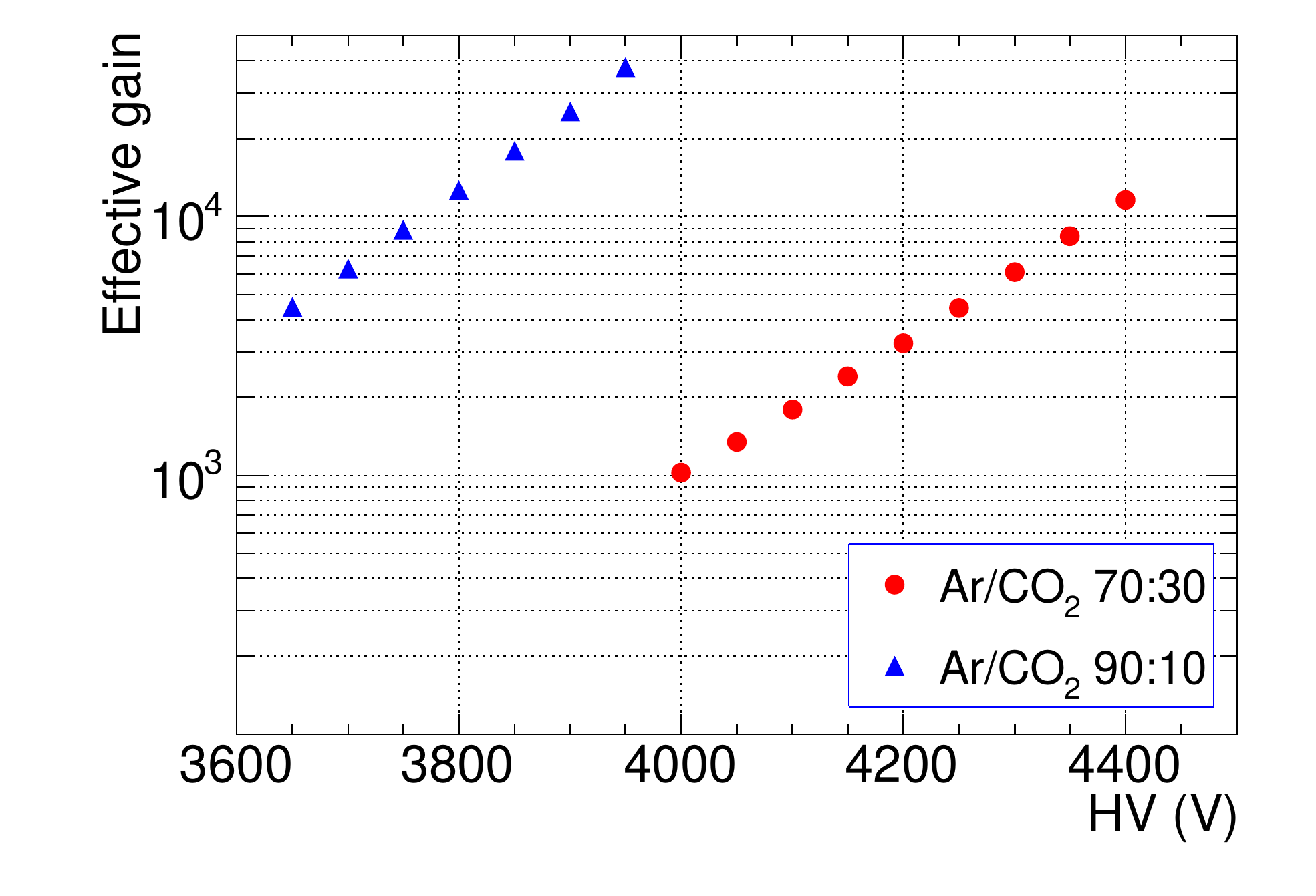}
\caption[short]{Effective gain of the GEM detector as a function of HV
  for the two gas mixtures.}
\label{gain}
\end{center}
\end{figure}

One of the important characteristics of any detector is its gain.
For this measurement, the detector was tested with a $^{55}$Fe X-ray
source. The pulse height spectrum for $^{55}$Fe has already been 
shown in Fig.~\ref{fe}. The gain is calculated with the formula:
\begin{equation}
G_{\rm eff} = \frac{M.K_{\rm elec}}{N_{\rm p} . q_{\rm e}}
\end{equation}
where $G_{\rm eff}$ is the effective gain of the detector, 
$M$ is the mean ADC value of the main peak of the $^{55}$Fe 
 spectrum, $K_{\rm elec}$ is the electronics gain factor, $N_{\rm p}$ is the number of primary electrons produced by full energy deposition of 5.9~keV X-ray in the drift volume and $q_{\rm e}$ is the electron charge. 
Fig.~\ref{gain} shows the effective gain as a function of HV for two
gas mixtures. As expected, gain increases with HV and has an 
exponential trend. For the 90:10 gas mixture, the gain curve is
shifted toward lower HV values with respect to the 70:30 gas mixture.

\begin{figure}[!ht]
\begin{center}
\includegraphics[scale = 0.4]{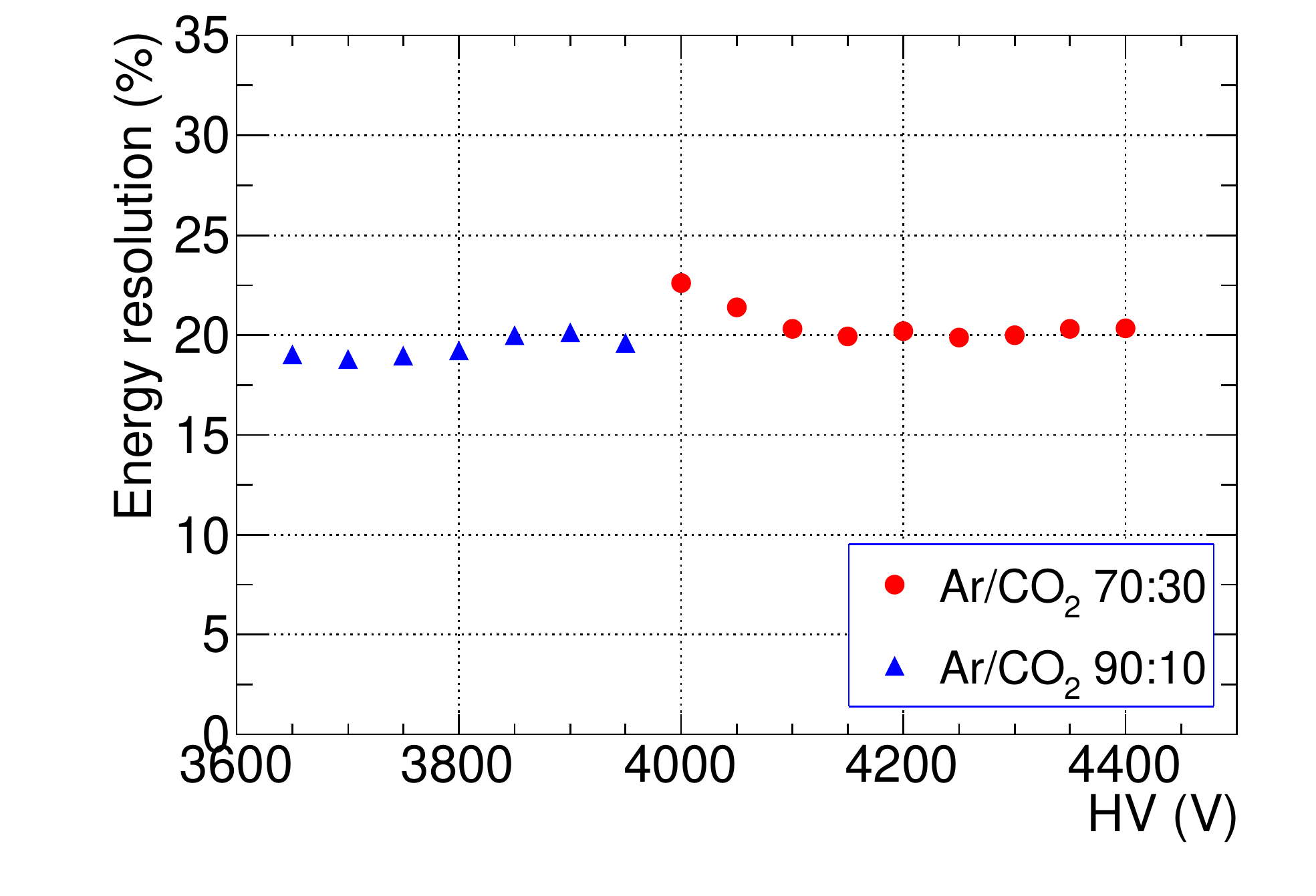}
\caption[short]{Energy resolution of the GEM detector as a function of
  applied HV for the two gas mixtures.}
\label{reso}
\end{center}
\end{figure}
\subsection{Energy resolution}
The energy resolution of the detector is calculated from the Gaussian fit
parameters of the main peak of the $^{55}$Fe spectrum. The energy resolution for the two gas mixtures as a function of HV is shown in Fig.~\ref{reso} in terms of FWHM. The energy resolution shows little variation over the range of voltage studied for both gas mixtures and is about 20\% (FWHM). 
\subsection{Time resolution}

Time resolution of GEM detectors is determined by the spread in
signal formation time for different events. Time
resolution depends on several factors, the most important being the
electron drift velocity. Drift velocity depends on the gas type and
the electric field value. In the present setup, 
time resolution with the Ar and CO$_2$ gas mixture in 70:30 proportion was measured using the $^{106}$Ru-Rh $\beta$-source. The three-fold scintillator
trigger described in the previous section was used as start signal 
and the signal from the GEM detector processed through a fast
amplifier was used as the stop signal.
The time difference between the start and the stop
signals was measured by an ORTEC 567 Time to Amplitude Converter
(TAC).
\begin{figure}[!ht]
\begin{center}
\includegraphics[scale = 0.4]{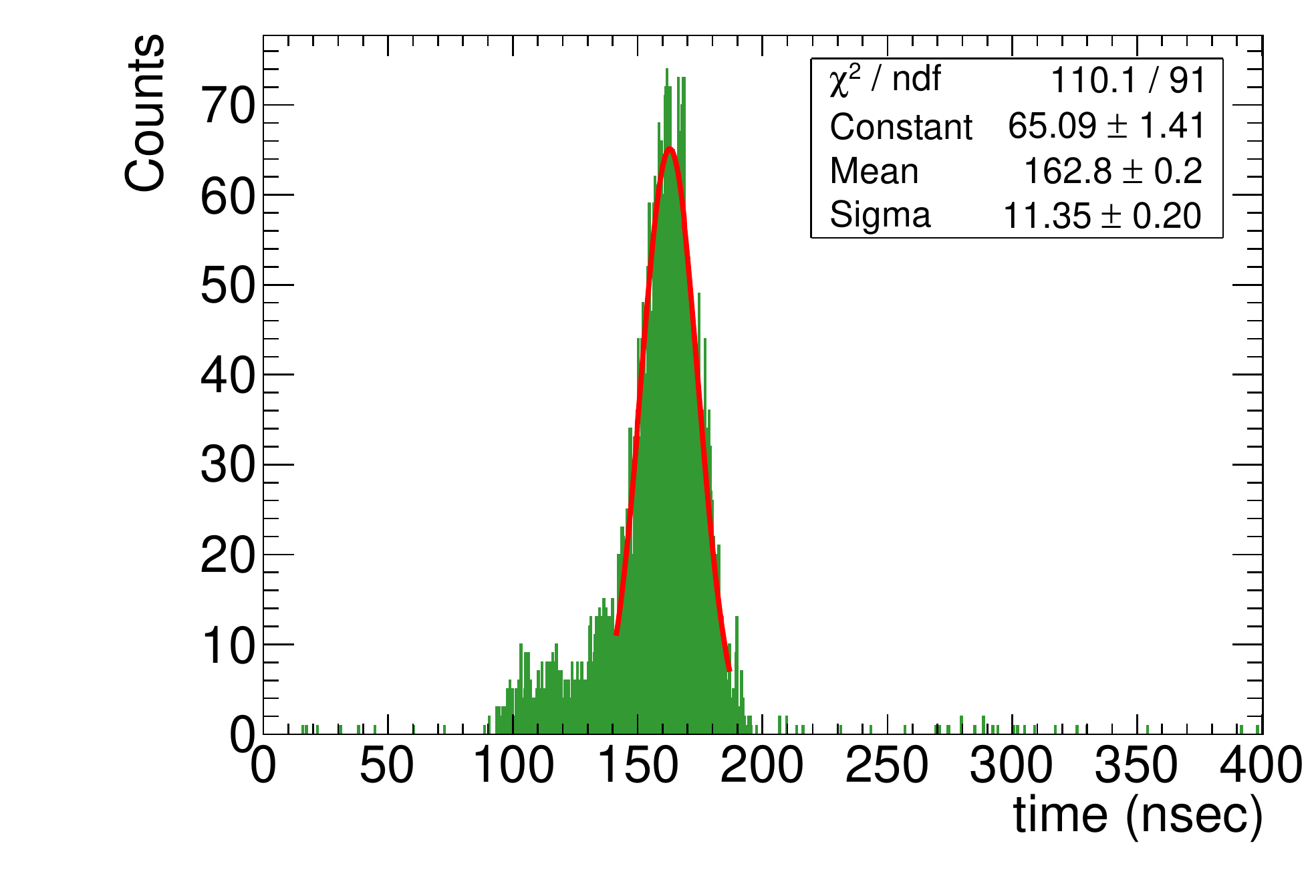}
\includegraphics[scale = 0.4]{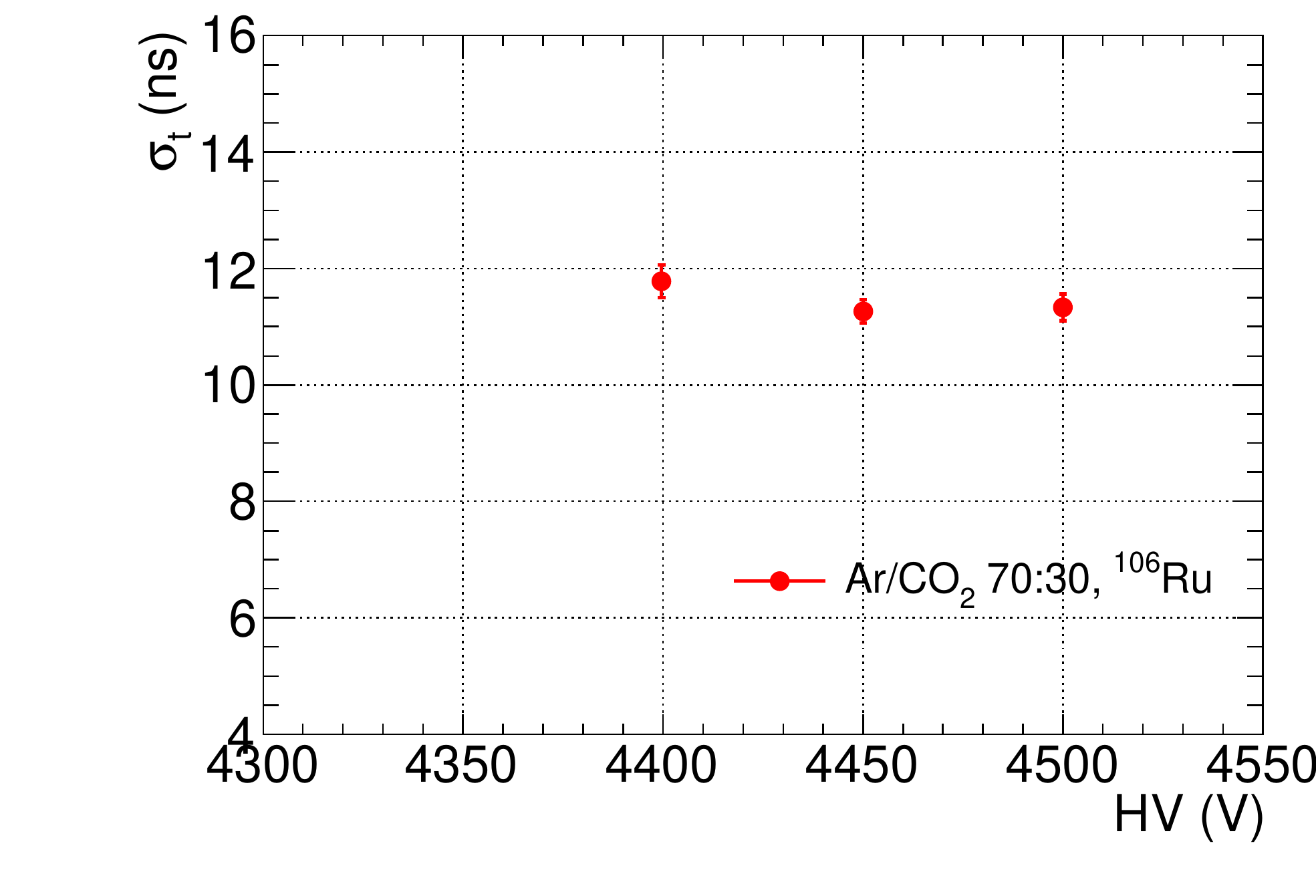}
\caption[short]{(top) Time spectrum at 4450 V and (bottom) time resolution as a function of applied HV.}
\label{time_res}
\end{center}
\end{figure}

The time spectrum and the plots for time resolution are shown in the top and bottom panels of Fig.~\ref{time_res},
respectively. The tail at the lower end of the time spectrum might be due to the fact that measurements were performed at high gain. Since the gain is high it might be possible to have a signal even if the primary ionisation happens to be in the first transfer gap. In that case the signal will be faster than the
majority of the signals whose primaries were generated in the drift
area. The results 
about the time resolution are in agreement with earlier results \cite{sauli2016,timing}. 
\section{Detector uniformity}
New generation high energy physics experiments require large area
detectors. The overall performance of these
detectors depends on gain uniformity, energy resolution and efficiency
over the entire active region. Several factors, like variations in
hole diameter, variations in gas gap due to inaccurate stretching,
etc., can lead to non-uniformity in the
detector~\cite{akl,qatar}. Thus it is essential to measure the
gain uniformity over the entire surface
area. In this work, a method has been used to measure the gain and energy resolution
in localised regions and then results obtained across different regions have
been compared.

\begin{figure}[!ht]
\begin{center}
\includegraphics[scale=0.14]{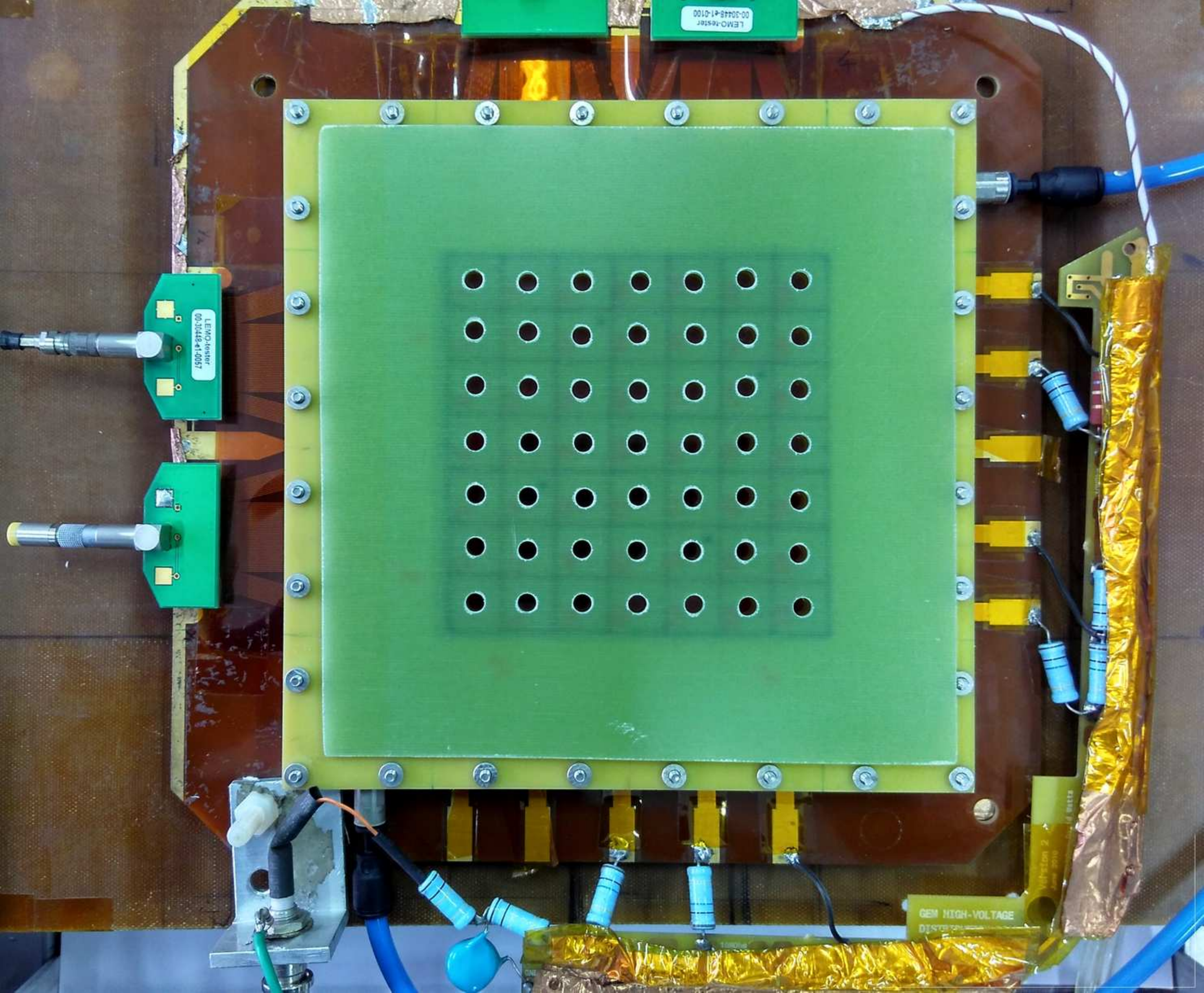}
\caption[short]{Test setup for gain and energy resolution uniformity.}
\label{uniform_setup}
\end{center}
\end{figure}

In the setup shown in Fig.~\ref{uniform_setup}, a thick
perforated PCB is placed above the GEM detector. 
The 10$\times$10 cm$^{2}$ central area of the PCB is divided
to 7$\times$7 zones of equal area. The $^{55}$Fe source was placed on
the centre of each zone and the resulting spectrum was recorded. For
each zone, sum-up signals from two associated readout connectors have
been read out. For the central zones, however, the charge spreads on
all four readout connectors. To have similar data taking conditions
for all the zones, it was decided to exclude the central
zones. In this process, the gain and energy resolution are measured
for 36 zones.

\begin{figure}[!ht]
	\begin{center}
		\includegraphics[scale=0.4]{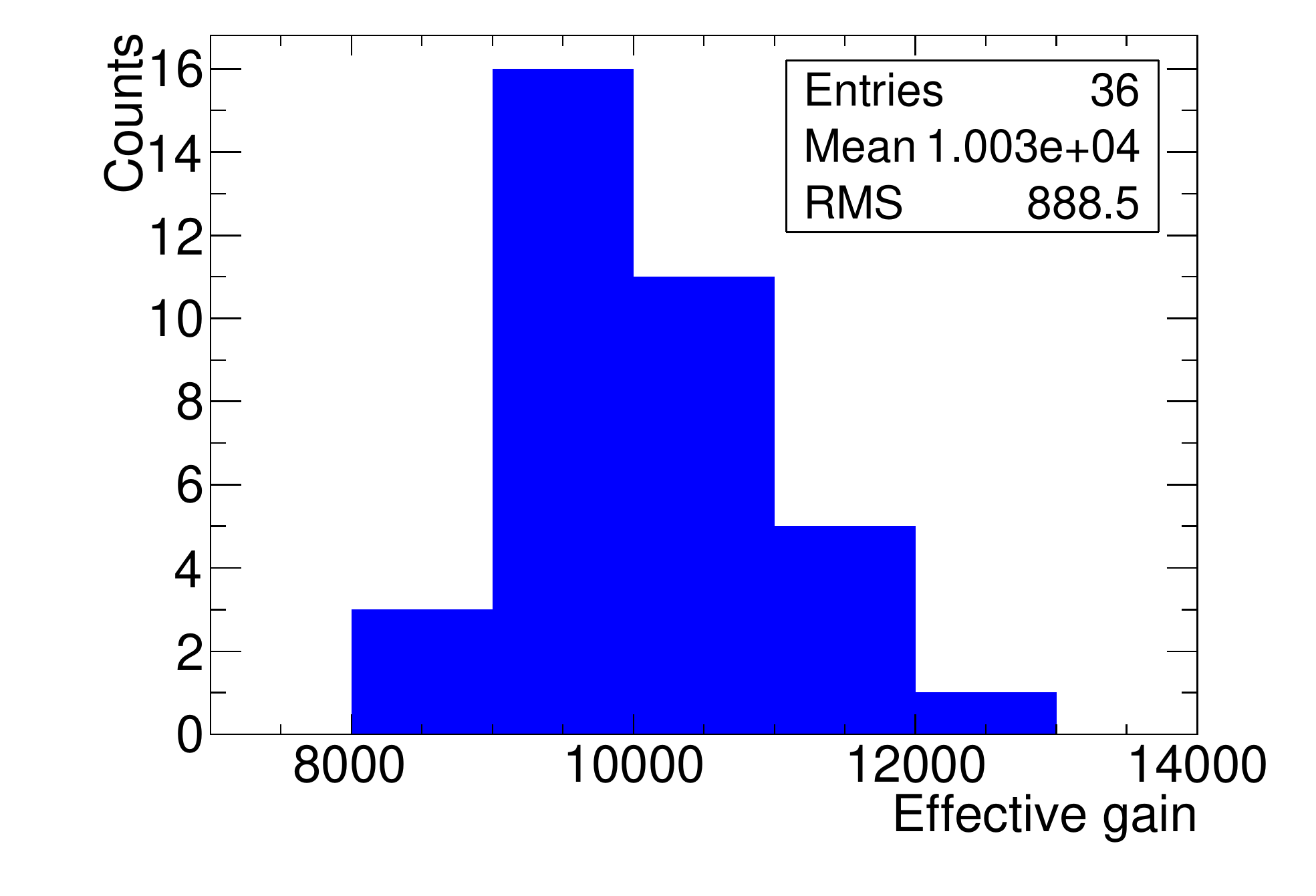}
		\caption[short]{Distribution of effective gain values
			for the 36 zones of the GEM detector at an applied
			voltage of 4400~V.}
		\label{uniformgain_1D}
	\end{center}
\end{figure}

\begin{figure}[!ht]
 \begin{center}
\includegraphics[scale=0.4]{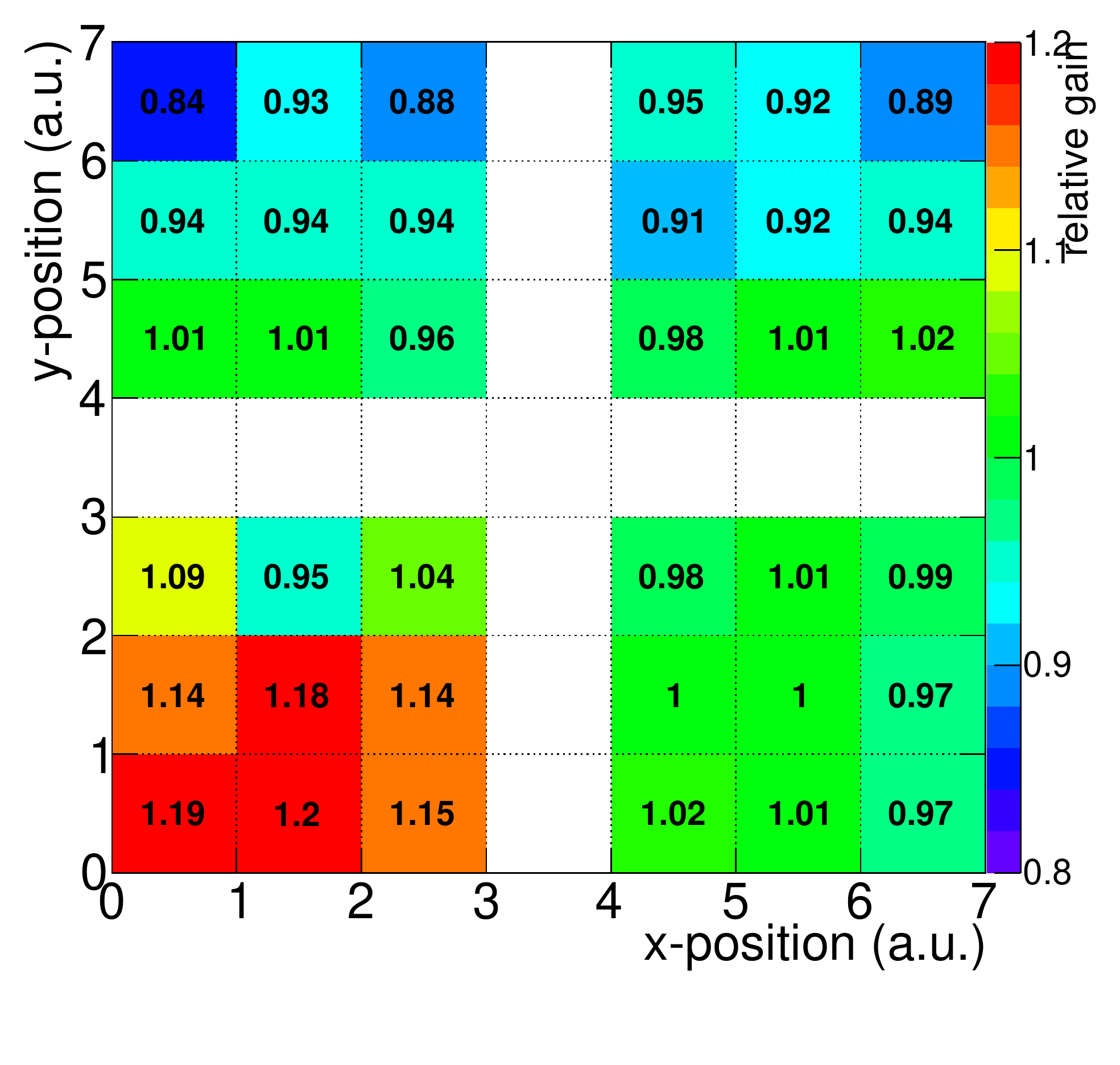}
\caption[short]{2D mapping of relative gain distribution in 36 zones
	of the GEM detector at an applied voltage of 4400~V.}
\label{uniformgain_2D}
 \end{center}
\end{figure}

\begin{figure}[!ht]
\begin{center}
	\includegraphics[scale=0.4]{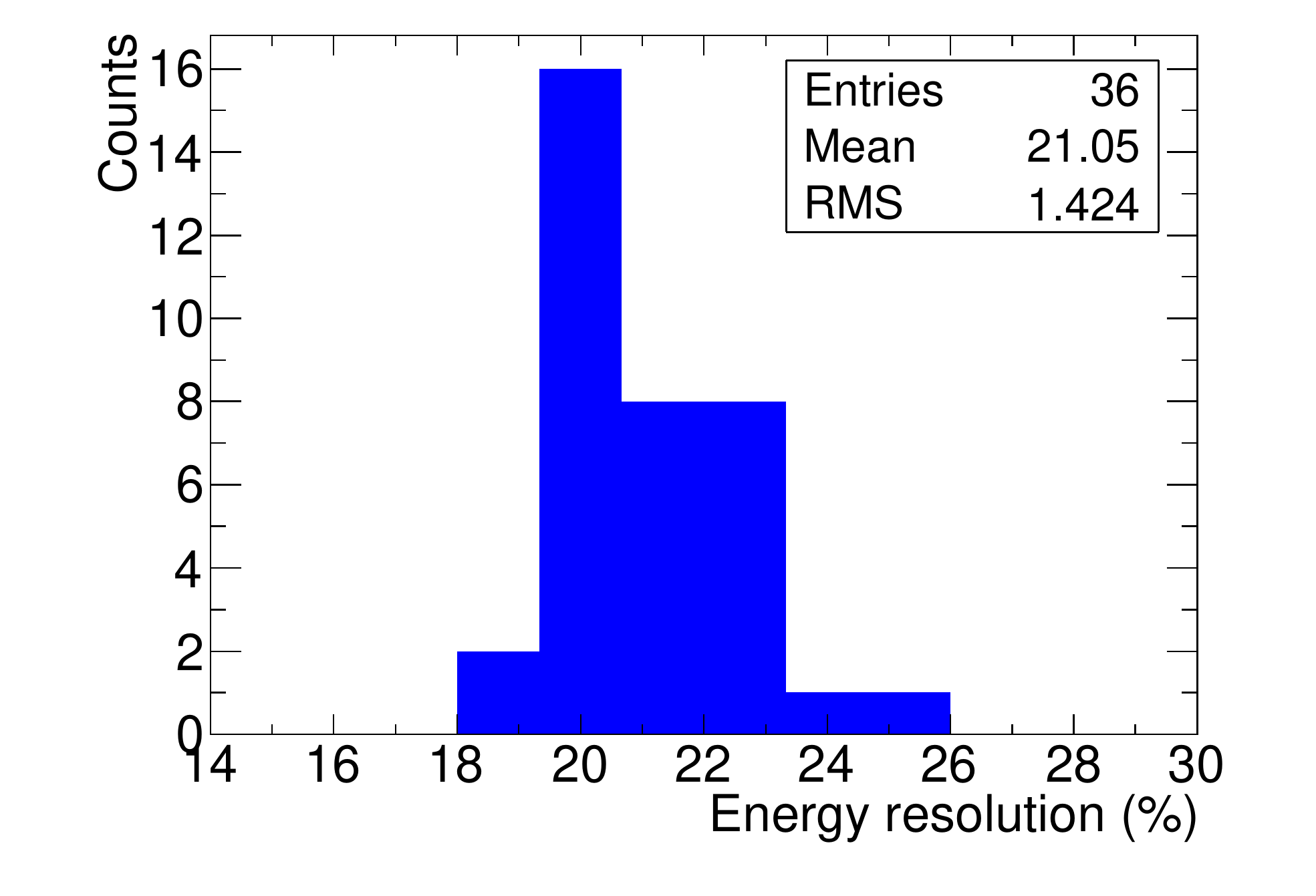}
		\caption[short]{Distribution of energy resolution
			values for the 36 zones of the GEM detector at an
			applied voltage of 4400~V.}
	\label{uniformreso_1D}
\end{center}
\end{figure}

Fig.~\ref{uniformgain_1D} shows the effective gain values for each
zone of the GEM detector at an applied voltage of~4400~V. The gain has
a mean value of 10030 at this voltage with a RMS of 8.8\%. As a
graphical representation, the relative gain of the detector measured
on each zone, has been plotted in Fig.~\ref{uniformgain_2D} after
normalising to the mean value. The gain uniformity result is in good
agreement with literature~\cite{simon}. 
The energy resolutions obtained for each of the 36 zones at
4400~V  is shown in Fig.~\ref{uniformreso_1D}.
The mean value of the energy resolution (FWHM) is 21\% with a
RMS of 6.7\%.

\section{Summary and outlook}
A detailed study of a 10$\times$10~cm$^2$ triple GEM
detector filled with a gas mixture of Ar+CO$_2$ of 70:30 and 90:10
proportions has been performed. 
Tests were conducted using cosmic rays trigger, a $^{106}$Ru-Rh
$\beta$-source and  a $^{55}$Fe X-rays source.
A plateau in the efficiency around $\sim$95\% has been obtained at 
different operating voltages for the two Ar+CO$_2$ gas mixtures. 
The energy resolution of the detector was measured to be around 20\%
for FWHM around the plateau region. A time resolution of $\sim$10~ns
has been achieved with the Ar/CO$_{2}$ 70:30 gas mixture.

Uniformity in gain and energy resolution of the detector have been studied by
dividing the detector in 7$\times$7 zones and observing the response
to a $^{55}$Fe source for each zone separately. The RMS variations of gain and energy resolution are 8.8\% and 6.7\%, respectively over the entire area. This gain fluctuations can be used in
simulations in order to quantify the overall detector response in
experiments. The method described in this paper of measuring uniformity in detector gain gives a quantitative account of these
parameters, and will prove to be very useful for quality
assurance checks for large area GEM detectors.

\bigskip

\noindent
{\bf Acknowledgment}
RNP acknowledges the receipt of UGC-NET fellowship and SB acknowledges the support of DST-SERB Ramanujan Fellowship. YPV thanks Indian National Science Academy, New Delhi for the Senior Scientist position. RNP acknowledges valuable discussions with Rob Veenhof and Dariusz Miskowiec.

\end{document}